\documentclass[pre, aps,twocolumn,floatfix,superscriptaddress]{revtex4-1}
\usepackage{amsmath,amssymb,graphicx,cases}
\usepackage{epsfig}
\usepackage{textcomp}
\usepackage{epstopdf}
\usepackage{float}
\usepackage{braket}
\usepackage[normalem]{ulem}
\usepackage[utf8]{inputenc}

\usepackage[normalem]{ulem}
\usepackage{dsfont}
\newcommand{\stkout}[1]{\ifmmode\text{\sout{\ensuremath{#1}}}\else\sout{#1}\fi}

\usepackage{fixmath}

\usepackage{color}
\definecolor{Blue}{rgb}{0.00, 0.00, 1.00}
\definecolor{Red}{rgb}{1.00, 0.00, 0.00}
\definecolor{Green}{rgb}{0.00, 0.60, 0.00}

\newcommand{\nn}{\nonumber}
\newcommand{\be}{\begin{equation}}
\newcommand{\ee}{\end{equation}}
\newcommand{\bea}{\begin{eqnarray}}
\newcommand{\eea}{\end{eqnarray}}

\usepackage{multirow}
\usepackage{float}
\usepackage{caption, subcaption}

\usepackage[colorlinks=true, urlcolor=blue, anchorcolor=blue, citecolor=blue,filecolor=blue,linkcolor=blue,menucolor=blue]{hyperref}

\begin{document}
\title{Dynamical phase transition in the occupation fraction statistics for non-crossing Brownian particles}

\author{Soheli Mukherjee}
\email{soheli.mukherjee2@gmail.com}

\affiliation{Department of Solar Energy and Environmental Physics,
  Blaustein Institutes for Desert Research, Ben-Gurion University of
  the Negev, Sede Boqer Campus, 8499000, Israel}

\author{Naftali R. Smith}
\email{naftalismith@gmail.com}

\affiliation{Department of Solar Energy and Environmental Physics,
  Blaustein Institutes for Desert Research, Ben-Gurion University of
  the Negev, Sede Boqer Campus, 8499000, Israel}


\begin{abstract}

We consider a system of $N$ non-crossing Brownian particles in one dimension.
We find the exact rate function that describes the long-time large deviation statistics of their occupation fraction in a finite interval in space.
Remarkably, we find that, for any general $N \geq 2$, the system undergoes $N-1$ dynamical phase transitions of second order. 
The $N-1$ transitions are the boundaries of $N$ phases that correspond to  different numbers of particles which are in the vicinity of the interval throughout the dynamics.
We achieve this by mapping the problem to that of finding the ground-state energy for $N$ noninteracting spinless fermions in a square-well potential. The phases correspond to different numbers of single-body bound states for the quantum problem.
We also study the process conditioned on a given occupation fraction and the large-$N$ limiting behavior.

\end{abstract}

\maketitle

\section{Introduction}

Fluctuations in stochastic systems are of central importance in statistical mechanics and other fields \cite{fluctuations, fluctuations1, OvMe2010, MeersonAssaf2017, neqpt, currentfluc, currentfluc1, condensation}. Examples include population dynamics  \cite{OvMe2010, MeersonAssaf2017}, non-equilibrium phase transitions \cite{neqpt}, current statistics \cite{currentfluc, currentfluc1}, condensation phenomena \cite{condensation}, etc to name a few. In particular large deviations (or rare events) have been a central theme of interest over the past few decades \cite{ellis, hugo2009, Varadhan, O1989, DZ, Hollander, Jack20}. The fluctuations of the quantities of interest are encoded in the large deviation functions (LDFs), which are analogous to thermodynamic potentials in equilibrium  systems \cite{hugo2009}. One of the most remarkable phenomena in the context of large deviations is dynamical phase transitions (DPTs), which correspond to singularities   (i.e., non-analyticities) in LDFs \cite{exclusion, exclusion1, baek, baek1, kafri, glass, singularities, NemotoEtAl19, CVC23}.

 The occupation (or residence) time  of a particle is the time that it spends in some specified spatial domain.
Fluctuations of the occupation time have been of interest in many random systems, including  coarsening  and  phase ordering dynamics in magnetic systems \cite{phaseorder, phaseorder1, phaseorder2}, financial time series \cite{finance}, random walks on graphs \cite{randomwalk, randomwalk1}. Since the seminal work of L\'{e}vy \cite{levy}, who computed the exact probability distribution of the occupation time  for an ordinary
Brownian motion, there have been studies of the fluctuations of occupation time for various processes  in or out of  equilibrium \cite{kac, kac2,   godreche, majumdar1,  TouchetteMinimalModel, Touchetteoneparticle, AKM19, KA22, BD15, BD09, BB11, BB05, BB07, SHB09, Barkai06}, because of their potential applications in many physical systems. For example,  to analyze the morphological dynamics of interfaces \cite{morphology}, the fluorescence intermittency emitting from colloidal semiconductor dots \cite{semidots}, theory and experiments of  blinking quantum dots \cite{SHB09}, optical imaging \cite{imaging} etc.

Recently, Tsobgni Nyawo and Touchette studied  fluctuations of occupation fraction in a closed interval for a Brownian motion with and without drift \cite{TouchetteMinimalModel, Touchetteoneparticle}. 
They demonstrated that this relatively simple model exhibits a DPT in the fluctuations of the occupation time in presence of a drift in the long-time limit.
Their analysis of this model used the Donsker-Varadhan (DV) large-deviation formalism \cite{hugo2009, donsker, donsker1, majumdar4, majumdar5, Jack20}, which maps the problem onto that of finding the ground-state energy of a quantum problem of a single particle inside  an effective potential well. 
The transition  occurs between the escape and confinement of the Brownian motion and is first order in nature. Similar DPTs have been observed in many systems in the limit of low noise and/or large system size \cite{exclusion, exclusion1, baek, baek1, kafri, glass, singularities, fluctuations1, lownoise, naftali2}. To observe the DPT in drifted Brownian motion occupation fraction it is sufficient to consider the long-time limit without taking any additional limits of small or large parameters \cite{TouchetteMinimalModel, Touchetteoneparticle}.

It is natural to ask how the occupation fraction fluctuates for $N>1$ Brownian particles, conditioned not to cross each other  (`vicious' Brownian motions). The non-crossing condition introduces correlations between the particles, making this a nontrivial, many-body problem.
Since the pioneering works \cite{genne} and \cite{fishersurvival}, non-crossing random walkers have been studied in the context of wetting and melting \cite{fishersurvival}, networks of polymers \cite{polymers},  persistence properties in nonequilibrium systems \cite{viciouswalker1} and more. 
Non-crossing Brownian motions are also known to be closely related to random matrix theory because the joint distribution of their positions (which has been studied in many contexts \cite{schehr1, schehr, schehr2, NIBM, GLMS19, fermions, GMS21}) coincides, in some special cases, to that of eigenvalues of certain types of random matrices \cite{Dyson62, rmt, rmt1}.

In this work, we extend the study of occupation fraction statistics to $N>1$ non-crossing Brownian particles in an interval $[-l,l]$ for a given time interval  $[0,T]$. By extending the DV formalism to non-crossing Brownian motions, we study the large deviation function for all $N \geq 1$. 
 We find that the DV formalism maps this problem to $N$ noninteracting, spinless \emph{fermions} in a square-well potential, cf. \cite{GLMS19, fermions}.
 Interestingly, we find that the large deviation functions show multiple singularities: The system undergoes  $N-1$ second order DPTs. 
 These transitions are of very different nature to the DPT found for a single Brownian particle in \cite{TouchetteMinimalModel, Touchetteoneparticle}.
 In particular, they occur in the absence of a drift, i.e., the dynamics obey time-reversal symmetry.
In each of the $N$ different phases, separated by the DPTs, a different number of particles remains in the vicinity of the interval for the entire dynamics, while the other particles wander away from the interval.

The paper is organized as follows. In Sec. \ref{sec2} we define the model and present the scaling behavior of the fluctuations of the occupation fraction. In Sec. \ref{sec3} we show how to extend the Donsker-Varadhan (DV) formalism to our system. In Sec. \ref{sec4}, we solve the DV problem and show that there are $N-1$ second order phase transitions. We give explicit results for $N=2$.
We also discuss the process conditioned on a given value of the occupation fraction, and some limiting behaviors that emerge in the limit $N\gg1$. 
Finally, we summarize and conclude in Sec. \ref{sec5}. Some details of the calculations are given in the Appendices.


\section{Model}\label{sec2}

We consider $N$ one-dimensional Brownian particles, which are defined by the following stochastic differential equations 
\be
\label{Langevin1}
\dot{X}_i(t)=\sqrt{2 D}\; \xi_i(t), \qquad i = 1,2,\dots,N,
\ee 
where $X_i$ is the position of the $i_{th}$ particle, $D$ is the diffusion constant and $\xi_i(t)$ are Gaussian white noises with $\langle \xi_i (t) \rangle =0$ and $\langle \xi_i (t) \xi_j (t') \rangle = \delta_{ij} \delta(t-t')$. 
Here $\langle . \rangle$ denotes the ensemble average over realizations of the noise.
We condition the particles to  be non-crossing, i.e,  $X_1(t) < X_2(t) < \dots < X_N(t)$ for all $t \in [0, T]$ \cite{footnote:noncrossing}.

The occupation fraction is defined as
\be
\label{rhoDef1}
\rho_T = \frac{1}{T} \int_0^T   \Big ( \mathds{1}_{[ -l,l ]}(X_1(t)) + \dots +   \mathds{1}_{[ -l,l ]}(X_N(t)) \Big) \, dt,
\ee
where $\mathds{1}_{[\ -l,l ]} (x)$ is the indicator function
\be
 \mathds{1}_{[\ -l,l ]} (x) =
 \begin{cases}
 1, & \text{for} \,\, x \in [-l,l],\\[2mm]
 0, & \text{otherwise.}
 \end{cases}
\ee
 $\rho_T$  is a random variable that takes values on $[0,N]$, where $\rho_T = 0$ ($\rho_T = N$)  corresponds to realizations in which none (all) of the Brownian particles stay in the interval $[-l,l]$ for the entire duration of the dynamics. For $T \longrightarrow \infty$, it becomes very unlikely for the Brownian particles to spend an extensive time $O(T)$ in the interval $[-l,l]$, so we expect that the probability distribution $P(\rho_T=\rho)$ concentrates around $\rho=0$.

 Rescaling space and time
\be
X/l\to X, \qquad 2Dt/l^{2}\to t,
\ee
Eqs.~\eqref{Langevin1} and \eqref{rhoDef1} become
\be
\dot{X}_i(t)=\xi_i(t), \qquad i = 1,2,\dots,N,
\ee 
and
\be
\rho_T = \frac{1}{T} \int_0^T   \Big ( \mathds{1}_{[ -1,1 ]}(X_1(t)) + \dots +   \mathds{1}_{[ -1,1]}(X_N(t)) \Big) \, dt,
\ee
with the rescaled $T$ (which equals $2DT/l^{2}$ in the original variables).
From this rescaling it follows that the distribution of $\rho_T$ takes the scaling form
\be
\mathcal{P}\left(\rho_T =\rho;D,l,T\right)=P\left(\rho_T =\rho;\frac{2DT}{l^{2}}\right)
\ee
in the original variables (where the dependence on the parameters is indicated explicitly, but will be suppressed for brevity below).

As we show below, in the long-$T$ limit, the fluctuations of $\rho_T$ follow the large deviation principle \cite{ellis, Varadhan, O1989, DZ, Hollander, hugo2009, Jack20, donsker, donsker1, gartner, ellis1},
which states that the probability distribution $P(\rho_T = \rho)$ scales as
\be 
P(\rho_T = \rho) \sim e^{-T I(\rho)}.
\ee
 This describes an exponential decay with $T$ at a rate described by the LDF or the rate function $I(\rho) \geq 0$, which is defined as
\be 
I(\rho) = - \lim_{T \longrightarrow \infty} \frac{1}{T} \ln P (\rho_T = \rho) 
\ee
 (In the physical variables, this translates to
$\mathcal{P}(\rho_T = \rho) \sim e^{-\left(2DT/l^{2}\right)I(\rho)}$
which is valid at 
$T\gg l^{2} / D$.)

In general, the rate function is difficult to calculate directly. However, according to the G\"{a}rtner-Ellis theorem \cite{gartner, ellis1} the rate function $I(\rho)$ can be calculated via the Legendre-Fenchel transform of the scaled cumulant generating function (SCGF), which is defined as \cite{hugo2009, gartner, ellis1} 
\be \label{eqscgf}
\lambda(k) = \lim_{T \longrightarrow \infty} \frac{1}{T} \,\, \ln \langle  e^{T k \rho_T} \rangle
\ee
The rate function is then expressed as the Legendre-Fenchel transformation of $\lambda(k)$
\be 
I(\rho) = \sup_{k} \lbrace  k \, \rho - \lambda(k) \rbrace
\ee
provided that $\lambda(k)$ exists and differentiable \cite{hugo2009}. If $\lambda(k)$ is convex, then the transform reduces to the Legendre transform, so that the conjugate variable of $k$,  $\rho$ is given by \cite{LegendreNutshell}
\be
\rho = \frac{d \lambda(k)}{d k} \, .
\ee
Hence, the problem reduces to that of calculating the SCGF of the occupation fraction of $N$ non-crossing Brownian particles. In the next section we explain how this can be done, by extending the DV formalism to noncrossing Brownian particles.


\section{Donsker-Varadhan formalism} \label{sec3}

An equivalent description of the dynamics of the $N$ Brownian particles is given in terms of the Fokker-Planck equation
\be
\frac{\partial Q(X_1, \dots , X_N, t)}{\partial t}= L^\dagger \,\,  Q(X_1, \dots, X_N, t) \, .
\ee
Here $Q(X_1, \dots, X_N, t)$ is the  unnormalized time dependent joint probability density function of the $X_i$'s,   $L^\dagger =  \frac{1}{2} \nabla^2$ is called the Fokker-Planck generator, with $\nabla^2 = \frac{\partial ^2}{ \partial X_1^2} + \dots + \frac{\partial^2}{\partial X_N^2} $ . $L^\dagger$ is a Hermitian  operator ($L^\dagger = L$) as the dynamics have time-reversal symmetry \cite{LegendreNutshell}. Due to the non-crossing condition, we consider only the domain $X_1 < \dots < X_N$, with the Dirichlet boundary conditions
 \be \label{eqbc}
 Q(X_1, \dots, X_N, t) \Big |_{X_i=X_{i+1}} \!\!\! = 0  \,\,\, \text{for any} \;  i  \! \in \! \lbrace 1, \dots, N-1 \rbrace.
 \ee
The \emph{normalized} joint probability density function of $X_1, \dots, X_N$ is
 \be 
P(X_1, \dots, X_N, t) = \frac{Q(X_1, \dots, X_N, t)}{\mathcal{N}(t)}
\ee 
where $\mathcal{N}(t)$ is the (time-dependent) normalization constant
\be 
\mathcal{N}(t) = \int_{-\infty}^{\infty} \!\! dX_1 \dots \int_{-\infty}^{\infty} \!\! dX_N Q(X_1, \dots, X_N, t).
\ee
At long times $t \longrightarrow \infty$,  $\mathcal{N}(t)$ goes as a power law  $\mathcal{N}(t) \sim t^{- N (N-1)/4}$  \cite{fishersurvival, fishersurvival1, viciouswalker, viciouswalker1, GLMS19}.

A mathematical framework for calculating large deviations of time-averaged quantities in stochastic processes of the form Eq.~\eqref{rhoDef1} (which are often referred to as dynamical or additive observables) was formulated by Donsker and Varadhan in \cite{donsker, donsker1}. 
 We will now extend this formalism to the case of noncrossing Brownian particles.
In order to calculate  $\lambda(k)$, the Feynman-Kac formula is used \cite{kac, majumdar5}, which states that the evolution of the generating function of $\rho_T$ involves a linear operator
\be
\label{hamiltonian}
 \mathcal{L}_k =  \frac{1}{2} \,\,  \nabla^2 +   k \Big (\mathds{1}_{[ -1,1 ]}(X_1) + \dots +   \mathds{1}_{[ -1,1 ]}(X_N) \Big ) \, ,
\ee
which is called the tilted generator.
The SCGF of $\rho_T$ is evaluated as  the largest eigenvalue of  $\mathcal{L}_k$  \cite{Touchette2018} (see Appendix \ref{feynman} for more detailed calculation)

 The eigenvalue problem  is defined by the equation
\be \label{eveq}
\mathcal{L}_{k} \,\, r_{k}(X_1, \dots , X_N)=\lambda(k) \,\, r_{k}(X_1, \dots , X_N)
\ee
where $r_k(X_1, \dots , X_N)$ is the right eigenfunction of $\mathcal{L}_k$,  together with the same boundary conditions as for $Q$, i.e., 
 \be \label{eqbcef}
 r(X_1, \dots , X_N, t) \Big |_{X_i=X_{i+1}}  \!\! = 0  \,\,\, \text{for any} \,\,\,  i \in  \lbrace 1, \dots, N-1 \rbrace .
 \ee
Note that as the operator $\mathcal{L}_k$ is Hermitian, its left eigenfunctions are the same as $r_k(X_1, \dots, X_N)$.
We are generally interested in normalizable solutions, and we (arbitrarily) choose the normalization constant to be 1
\be 
\int_{-\infty}^{\infty} \! dX_1 \dots \int_{-\infty}^{\infty} \! dX_N \,  |r_k(X_1, \dots , X_N)|^2 = 1 \, .
\ee

\begin{figure}
\includegraphics[width=0.98\linewidth,clip=]{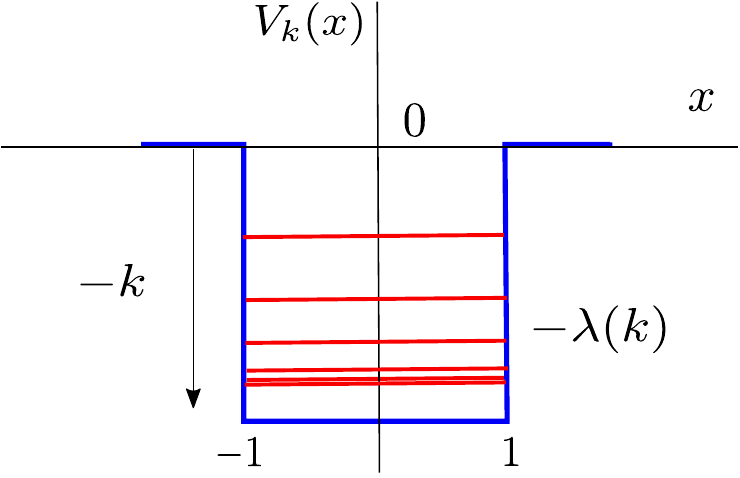}
\caption{(Color online)  The effective quantum problem of $N$ fermions in a square potential well \eqref{Vkdef} of depth $-k$, whose solution yields the behavior of fluctuations of the occupation fraction in the interval $[-1, 1]$ of $N$   non-crossing Brownian particles. The solid red lines represent the  single-particle energy levels  occupied by the fermions.  $- \lambda(k)$ is the many-body ground-state energy.}
\label{schematic}
\end{figure}

Eq.~\eqref{eveq} with the tilted operator \eqref{hamiltonian} and the boundary condition \eqref{eqbcef} can be recognized as the time-independent Schr\"{o}dinger equation of $N$ noninteracting, spinless fermions up to an overall minus sign with  $\frac{\hslash^2}{2 m} = \frac{1}{2} $ in a finite potential well 
\be
\label{Vkdef}
V_k(x) = - k \mathds{1}_{[-1,1]}(x) \, .
\ee 
Fig. \ref{schematic} shows a schematic plot of the equivalent quantum problem. 
This implies that the $\lambda(k)$ equals minus the (many-body) ground state energy of $- \mathcal{L}_k$.  For $N=1$ the problem reduces to that of a single particle in the potential well $V_k$,   which was solved in \cite{TouchetteMinimalModel, Touchetteoneparticle}.


\section{Results} \label{sec4}

The SCGF $\lambda(k)$ for general $N$ is calculated from the ground state energy of $N$ noninteracting, spinless fermions in a square well potential of depth $k$, which is given by the sum of the single-body energy levels.
For a single Brownian particle ($N=1$), there is always a bound state for any $k > 0$ given by cosine inside the well and two decaying exponentials outside the well. The single-body energy levels of a quantum particle inside a square-well potential 
can be found in any standard textbook of quantum physics \cite{hall, qm}, and each of them is given by the solution to one of two transcendental equations (see below). For any general $N$, an interesting effect occurs:
As $k$ is increased, the number of solutions to the transcendental equations increases, and, as shown below, this results in $N-1$ dynamical phase transitions which correspond to critical values of $k$.

 In this section, we first briefly recall the  single-body energy levels of a quantum particle in a finite potential well. Then, we give a brief review of the results for the case $N=1$ \cite{TouchetteMinimalModel, Touchetteoneparticle}. 
 Next, we study the problem for general $N$, giving explicit results for the case $N=2$ which is already sufficient to observe the dynamical phase transition. 
 We then consider the large-$N$ limit where we uncover a universal behaviour of the system. We also discuss the conditioned process, both for general $N$ and in the large-$N$ limit.



\subsection{Single-body energy levels}

Due to the symmetry in the potential Eq. \eqref{Vkdef}, the single-body wave functions of bound states are either symmetric or antisymmetric. For any $k>0$, there always exists at least one bound state with a symmetric wave function \cite{hall}. As the depth of the potential $k$ increases, the number of bound energy levels increases, with the wave functions alternating between symmetric and antisymmetric. The energy levels $-\lambda_1, -\lambda_2, \dots$ satisfy the following transcendental equations \cite{hall, qm}
\be \label{eq3}
A+A\tan^{2}\left(\sqrt{2A}\right)=k \qquad  \text{for symmetric case}
\ee
and
\be \label{eq3new}
A+A\cot^{2}\left(\sqrt{2A}\right)=k  \qquad  \text{for antisymmetric case}
\ee
where
\be \label{eq31new}
A=k-\lambda_i(k).
\ee
The $i^{\text{th}}$ energy level ($i=1,2,\dots$) is the solution of Eq.~\eqref{eq3} (for odd $i$) or Eq.~\eqref{eq3new} (for even $i$)
for the following range of $A$'s
\be \label{eqA}
\frac{\pi^{2}}{8} (i-1)^2<A_{i}< \frac{\pi^{2}}{8} i^2 \, .
\ee

\subsection{$ N=1$}

For a single ($N=1$) Brownian motion the problem was solved in \cite{TouchetteMinimalModel, Touchetteoneparticle}. The calculation of $\lambda(k) = \lambda_1(k)$ boils down to calculating the ground state energy of one quantum particle trapped in a potential well with depth $k$. The ground state wave function is symmetric satisfying Eq. \eqref{eq3}, where $0 \leq k < \infty$ and $0 \leq A \leq \frac{\pi^2}{8}$.
Eqs.~\eqref{eq3} and \eqref{eq31new} thus give $\lambda(k)$ in a parametric form (with $k$ and $\lambda$ both given explicitly as functions of the parameter $A$).

The conjugate variable $\rho$ and the rate function $I(\rho)$ of the problem can also be expressed in terms of the variable $A$, thus giving the rate function in a parametric form:
\bea
\rho &=& \frac{d\lambda}{dk}=\frac{d\lambda/dA}{dk/dA} = 1-\frac{\cos^{2}\left(\sqrt{2A}\right)}{1+\sqrt{2A}\tan\left(\sqrt{2A}\right)} \, , \\
I &=& \rho \,  k - \lambda = \frac{A\sqrt{2A}\tan\left(\sqrt{2A}\right)}{\sqrt{2A}\tan\left(\sqrt{2A}\right)+1} \, .
\eea

  The asymptotic behaviours of $\lambda(k)$ and $I(\rho)$ are given by \cite{Touchetteoneparticle} (see also Appendix \ref{assymp calculation} for a derivation)
\bea
\label{inftyasymscgf}
\lambda\left(k\right) &\simeq& \begin{cases}
 2k^{2}, & k \longrightarrow 0,  \\[1mm]
 k-\frac{\pi^{2}}{8}+\frac{\pi^{2}}{4\sqrt{2k}}, & k \longrightarrow \infty,
\end{cases} \\[1mm]
\label{inftyasymrate2}
I\left(\rho\right) &\simeq&
\begin{cases}
\frac{\rho^{2}}{8}, & \rho  \longrightarrow 0,  \\[1mm]
  \frac{\pi^{2}}{8}-3\left(\frac{\pi^{2}}{8\sqrt{2}}\right)^{2/3}\left(1-\rho\right)^{1/3}, & \rho \longrightarrow 1.
\end{cases}
\eea

\begin{figure}
\includegraphics[width=0.98\linewidth,clip=]{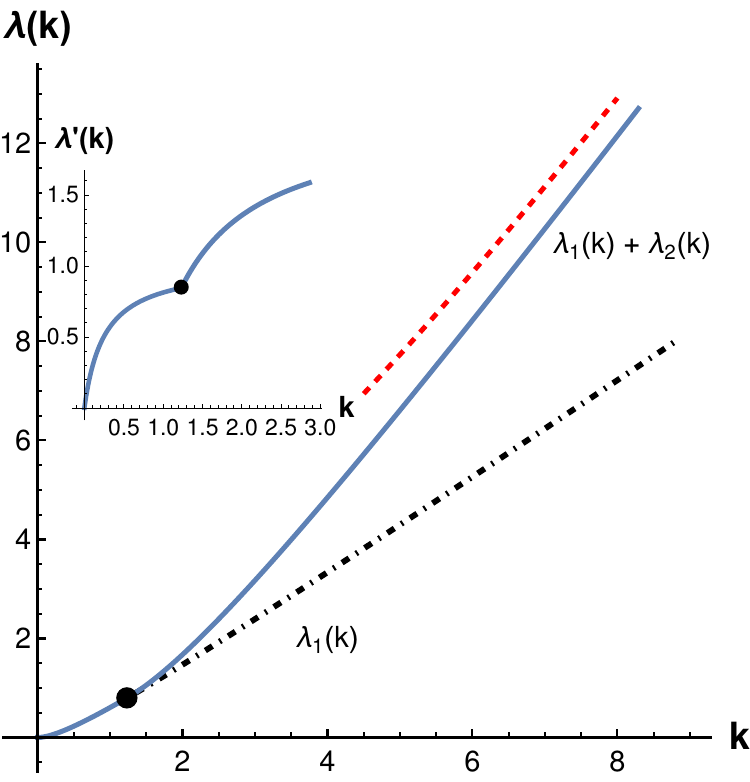}
\caption{(Color online) Solid line: The plot of the  exact 
SCGF $\lambda(k)$ as a function of $k$ for $N=2$ non-crossing Brownian particle occupation fraction. The solid circle represents the critical point $k_c = \pi^2 / 8$. For $k<k_c$,  $\lambda(k)$ coincides with the SCGF $\lambda_1(k)$ for $N=1$, and for $k > k_c$, the SCGF is $\lambda(k)=\lambda_1(k)+ \lambda_2(k)$ where $-\lambda_2(k)$ is the energy of the first excited state for the potential \eqref{Vkdef}. The dot-dashed line represents the continuation of the single particle SCGF for $k > k_c$. The red dashed  line depicts the asymptotic behavior of $\lambda(k)$ for $k \gg 1$. 
The second derivative of $\lambda(k)$ has a jump discontinuity at $k=k_c$. This can be seen as a corner singularity in the inset which shows $d\lambda/dk$ as a function of $k$. 
}
\label{fig2}
\end{figure}

The SCGF and rate function for $N=1$ are shown in Figs.~\ref{fig2} and \ref{fig3} respectively ($\lambda_1(k)$ and $I_1(\rho)$ respectively).

\subsection{General $N $}

For $N>1$, the problem is mapped to $N$ noninteracting, spinless fermions in the potential well \eqref{Vkdef}. As $k$ is increased, the number of single-body bound states increases and the fermions alternately occupy the odd and even energy levels which are found from Eqs.~\eqref{eq3} and \eqref{eq3new}, respectively. These energy levels correspond to multiple values of $A$ with ranges given by Eq. \eqref{eqA}.

Importantly, there are $N-1$ critical values of $k = k_{c,i} \equiv i^2 \frac{\pi^2}{8}$ for $i \in \lbrace 1, \dots, N-1\rbrace$,  at which the number of bound states increases. The critical values correspond to singularities of $\lambda(k)$ 
and these correspond to dynamical phase transitions, i.e., singularities of $I(\rho)$. Indeed, at $k$ slightly larger than $k_{c,i}$, $\lambda_{i+1}(k)$ behaves asymptotically as (for details see Appendix \ref{assymp calculation})
\be \label{asymscgfgeneral}
\lambda_{i+1}(k) \simeq  \frac{1}{2}\left(k-k_{c,i}\right)^{2}
\ee
and as a result, the SCGF $\lambda(k)$ shows a jump in the second derivative at $k=k_{c,i}$. As we show below, this leads to a jump in the second derivative of $I(\rho)$ which we interpret as a second order dynamical phase transition (see next subsection for details for the particular case $N=2$).

For $k < k_{c, N-1}$, $\lambda(k) = \lambda_1(k) + \dots + \lambda_M(k)$, where $M$ is the number of single-body bound states.  The eigenfunction $r_k(X_1, \dots, X_N)$ in Eq. \eqref{eveq}  is now given by the following Slater determinant \cite{footnote:SlaterNormalization}
\be
\label{SlaterDet}
\Psi\left(X_{1},\dots,X_{N}\right)=\left|\begin{array}{ccc}
\psi_{1}\left(X_{1}\right) & \dots & \psi_{N}\left(X_{1}\right)\\
\vdots & \ddots\\
\psi_{1}\left(X_{N}\right) &  & \psi_{N}\left(X_{N}\right)
\end{array}\right|
\ee
where $\psi_j(X_i)$ is the single particle state function. 
For $k < k_{c, N-1}$, $M<N$ so the many-body energy spectrum is continuous,  and $N-M$ of the fermions will not be localized around the interval $[-1,1]$.
As we show below, the interpretation of this in terms of the original problem is that $M$ of the Brownian particles remain in the vicinity of the interval $[-1,1]$ and the remaining $N-M$ particles do not.

In the next subsection, we will analyze the results for a fixed $N$. We consider  $N=2$ non-crossing Brownian particles occupation fraction, as it displays all of the interesting features for any general $N$.

\subsection{$N=2$}

For $N=2$, the two lowest single-body energy levels are found from Eqs.~\eqref{eq3}  and Eq. \eqref{eq3new} which together take the form
\be 
k=A_{2}\left(1+\cot^{2}\sqrt{2A_{2}}\right) = A_{1}\left(1+\tan^{2}\sqrt{2A_{1}}\right)
\ee
 There is a critical value of $k=k_c \equiv k_{c,1}= \frac{\pi^2}{8}$ (corresponds to $A_1 = 0.436191\dots$ and $A_2 = \frac{\pi^2}{8}$), below which only one physically possible solution $A_1$ contributes.

Whereas for $k > k_c$, the $\lambda_2(k)$ (corresponding to the solution $A_2$) contributes and the total SCGF of the problem is
\be 
\lambda(k)=\lambda_{1}(k)+\lambda_{2}(k).
\ee
Fig.~\ref{fig2} shows the plot of the SCGF for the two non-crossing Brownian particle occupation fraction (shown by the solid line). For $k < k_c$ the SCGF coincides with $\lambda_1(k)$. At $k=k_c$,  $\lambda_2(k)$ starts to contribute and continues to exist with  $\lambda_1(k)$ for all $k> k_c$.

\begin{figure}
\includegraphics[width=0.98\linewidth,clip=]{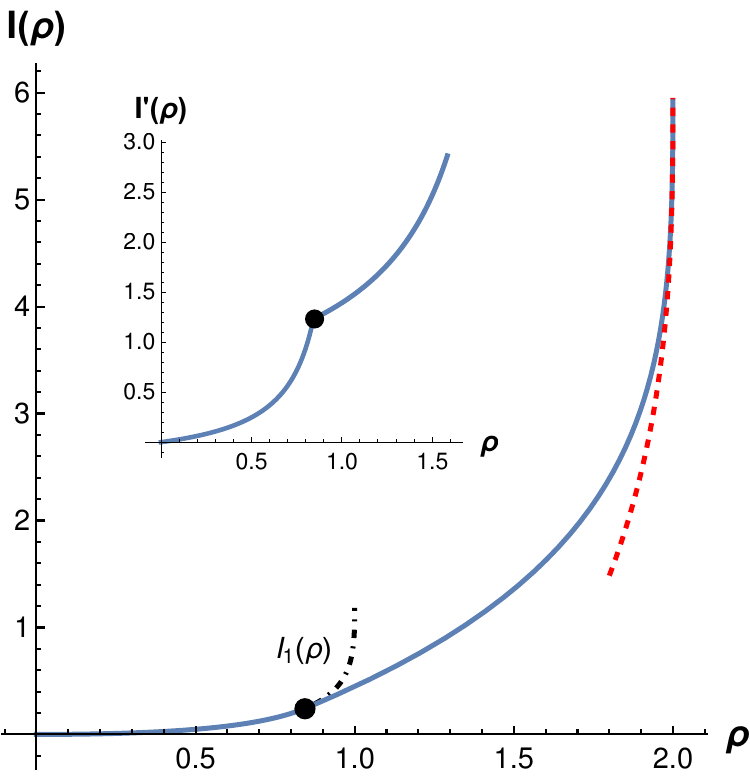}
\caption{ (Color online) Solid line:  The plot of the  exact rate function $I(\rho)$ as a function of the occupation fraction $\rho$ for $N=2$ non-crossing Brownian particles. The solid circle represents the critical point $\rho_c = 0.84375\dots$.  For $\rho < \rho_c$ the rate function coincides with the rate function for the corresponding single particle problem ($N=1$).
The dot-dashed line represents the continuation of the single particle rate function for $\rho > \rho_c$. The red dashed line depicts the asymptotic behavior of $I(\rho)$ for  $\rho \longrightarrow 2$.  The second derivative of $I(\rho)$ has a jump discontinuity at $\rho = \rho_c$, which we interpret as a second order dynamical phase transition. This can be seen as a corner singularity in the inset which shows $dI/d\rho$. 
}
\label{fig3}
\end{figure}

Similarly, for the rate function, there is a critical value of the occupation fraction $\rho = \rho_c = \lambda_1'(k_c) = 0.84375\dots$, below which the rate function coincides with the rate function $I_1(\rho)$ of the case $N=1$. For $\rho > \rho_c$ the rate function is given by the Legendre transform of the SCGF $\lambda_1(k) + \lambda_2(k)$. Fig.~\ref{fig3} shows the plot of the rate function for $N=2$.

Next, we  study the asymptotic behaviors of the SCGF, with particular emphasis on the behavior near the critical point $k \longrightarrow k_c$ ($\rho \longrightarrow \rho_c$) where we study the dynamical phase transition in detail.

In the $k \longrightarrow 0$ limit which corresponds to $\rho \longrightarrow 0$, the SCGF and the rate function coincide with their counterparts for $N=1$, and their asymptotic behaviors in these limits are given by the first lines of  Eqs.~\eqref{inftyasymscgf} and \eqref{inftyasymrate2}, respectively.

In the critical regime, the SCGF behaves as
\be \label{eqscgfcritical}
\lambda\left(k\right) \simeq \begin{cases}
\lambda_{1}\left(k\right), & k<k_{c},\\[2mm]
\lambda_1\left(k\right) + \frac{1}{2}\left(k-k_{c}\right)^{2}, & k - k_{c} \ll k_{c}.
\end{cases}
\ee
Similarly the rate function near $\rho_c$ behaves as
\be \label{eqratefunccritical}
\! I(\rho) \! \simeq \! \begin{cases}
\!\!I_1(\rho), & \rho<\rho_{c}, \\[2mm]
    \!\!I_{1}(\rho_{c})+k_{c}(\rho-\rho_{c})+ B \! \left(\rho-\rho_{c}\right)^{2}\!,   & \rho - \rho_{c} \! \ll \! \rho_{c}.
\end{cases}
\ee
where $B= \frac{1}{2\left[\lambda_{1}''(k_c)+1\right]} \simeq 0.447705\dots$. The detailed calculations are presented in Appendix \ref{assymp calculation}.

Clearly the second derivative of the SCGF jumps at the critical point. As we now show, the same is true for the rate function.
The first derivative of the rate function   $I(\rho)$ is continuous because $I_1'(\rho_c) = k_c$ but $I'(\rho)$ shows a corner singularity as shown in the bottom inset of Fig. \ref{fig3}. Indeed, the second derivative shows a jump discontinuity  (see Appendix \ref{assymp calculation})
\be 
I''(\rho)\simeq \begin{cases}
I_1''(\rho) = \frac{1}{\lambda_{1}''(k)} & \rho<\rho_{c} \\
 2 B    & \rho - \rho_{c} \ll \rho_{c}
\end{cases}
\ee
We interpret the jump in the 2nd derivative of $I$ as  a dynamical phase transition of second order \cite{baek, baek1}. 

In the $k \longrightarrow \infty$ limit the SCGF behaves as 
\be \label{inftyasymscgf2}
\lambda(k) \simeq 2k-\frac{5\pi^{2}}{8}+\frac{5\pi^{2}}{4}\frac{1}{\sqrt{2k}} \, .
\ee
Taking the Legendre transform of this expression, we find that the corresponding limit $\rho \longrightarrow 2$ of the rate function behaves as
\be \label{inftyasympI}
I(\rho) \simeq \frac{5\pi^{2}}{8}-3\left(\frac{5\pi^{2}}{8\sqrt{2}}\right)^{2/3}\left(2-\rho\right)^{1/3}
\ee
The details of the calculations are given in  Appendix \ref{assymp calculation}.
The asymptotic behaviors of the SCGF and the rate function are shown in Figs.~\ref{fig2} and  \ref{fig3} respectively by the dashed red lines.

\subsection{Large $N$ behavior}

\begin{figure*}
\centering
     \begin{subfigure}[b]{0.45\textwidth}
     \centering
         \includegraphics[width=\textwidth]{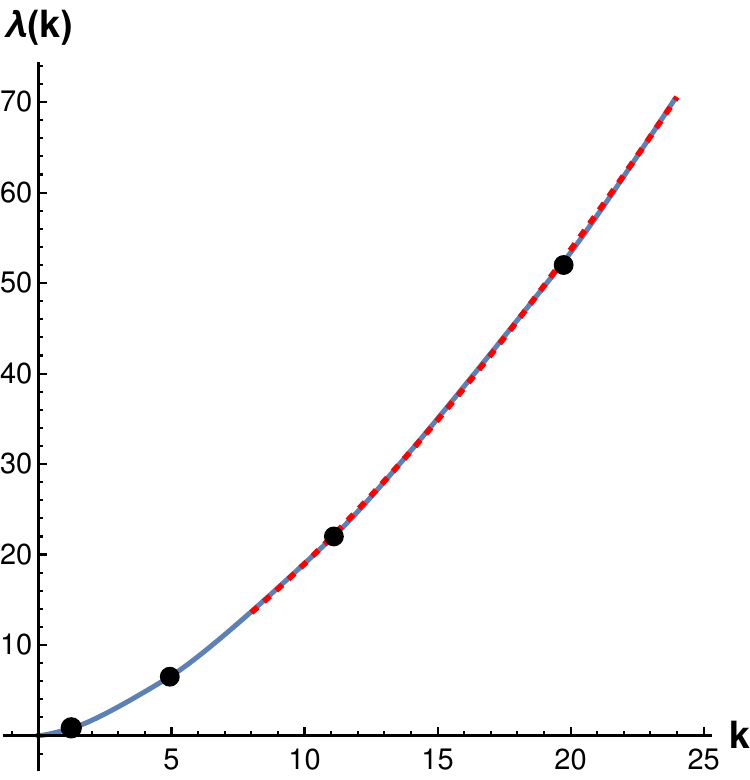}
         \caption{}
         \label{fig4}
     \end{subfigure}
     \hfill
     \centering
     \begin{subfigure}[b]{0.45\textwidth}
     \centering
         \includegraphics[width=\textwidth]{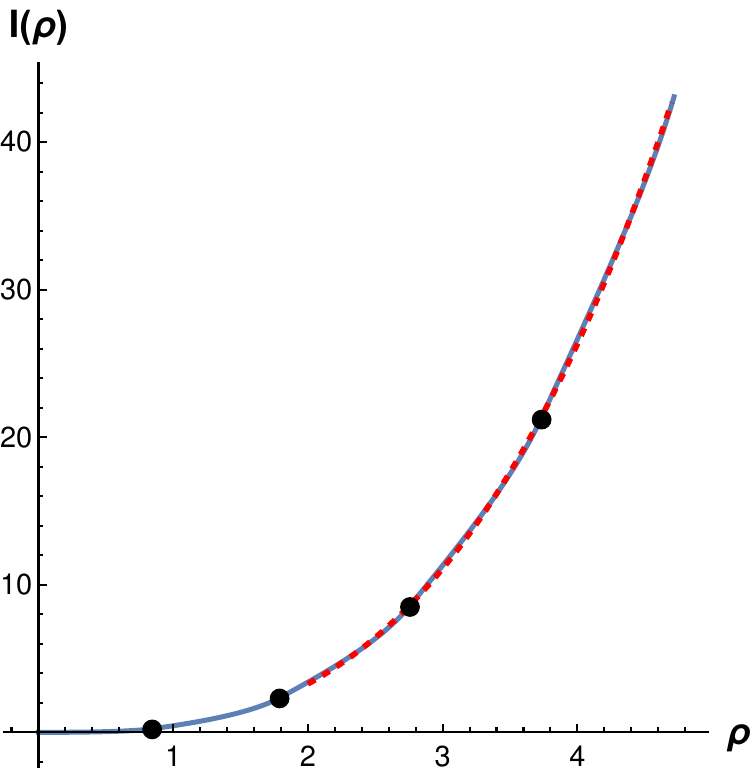}
         \caption{}
         \label{fig5}
     \end{subfigure}
     \hfill
        \caption{ The plot of the \textbf{(a)} SCGF as a function of $k$ and \textbf{(b)} rate function as a function of $\rho$ for $N=5$ Brownian particles. There are four critical points where the second-order dynamical phase transitions happen, shown by black dots. The red dashed lines depict the asymptotic large-$k$ and large-$\rho$ behaviours, see Eqs.~\eqref{scgfkLarge} and \eqref{IrhoLarge}. The exact and asymptotic expressions agree surprisingly well blue and are indistinguishable at the scale of the figure.}
        \label{fig6}
\end{figure*}

Let us consider the large-$N$ limit, where, as we show below, a universal behavior emerges at $\rho \gg 1$.
This corresponds to $k \longrightarrow \infty$, where the single-body energy levels can be approximated as the energy levels of particle-in-a-box i.e.  $E_i \simeq -k + \frac{\pi^2}{8} i^2$. $-\lambda(k)$ for large $N$ limit is thus the sum of the energy levels up to $M$, where $M$ is the energy level at which $E_M \simeq 0$, which gives
\be \label{approxM}
M \simeq \frac{2 \sqrt{2 k}}{\pi}
\ee
Thus the SCGF can be written as 
\bea \label{scgfkLarge}
\lambda(k) &=& - (E_1 + E_2 + ....+ E_M) \nn \\
&\simeq & M k -  \frac{\pi^2}{24} M^3
\simeq \frac{4 \sqrt{2}}{3 \pi} \,\, k^{3/2} \, .
\eea
Taking the Legendre transform, we obtain
\be 
\label{IrhoLarge}
I(\rho) \simeq \frac{\pi^2}{24} \,\, \rho^3
\ee
Eq.~\eqref{IrhoLarge} is valid in the limit $1 \ll \rho < N$. Despite the fact that $I(\rho)$ has $N-1$ singularities, we find that the limiting behavior \eqref{IrhoLarge} is nevertheless smooth.
Fig. \ref{fig6} shows the exact SCGF and rate function together with the limiting behaviors for  $N=5$, which turns out to be large enough to observe excellent agreement. There are four critical points at which the dynamical phase transitions of second order occur.

\subsection{Conditioned process}

$|\Psi(X_1, \dots, X_N)|^2$ is the joint distribution of the positions of the particles at some intermediate time $0<t<T$ conditioned on a given value of $\rho$ \cite{Touchetteoneparticle}.
Therefore, the $N$ phases correspond to conditioned processes in which $M$ particles stay in the vicinity of the interval $[-1,1]$ (where $M=1,\dots,N$). It is then intriguing to study the properties of the conditioned process in more detail. This can be done by exploiting the interpretation of the joint distribution as that of the positions of $M$ noninteracting fermions confined by the potential $V_k(x)$. 
 The latter system is closely related to that which was studied in Ref.~\cite{DLMSS21}. Below, we reiterate some of the main properties, but we refer the reader to Ref.~\cite{DLMSS21} for details.

Quantum fluctuations of interacting or noninteracting fermions, possibly confined by an external potential, have attracted much recent attention \cite{castin,CMV2012,eisler_prl,marino_prl,DLMS16PRA}, in particular due to their importance in applications such as experiments in cold atoms \cite{Fermicro1,Fermicro2,Fermicro3,Pauli,BDZ08,flattrap}.
Using a variety of analytical methods, such as determinantal processes and surprising connections to random matrix theory \cite{eisler_prl,marino_prl,DLMS16PRA,Calabrese_RMT,DLMS2019review, DLMSS21, SLMS21}, fundamental properties such as the spatial density of fermions and its correlations have been studied.
For $M$ noninteracting fermions in the presence of a trapping potential at zero temperature (i.e., in the many-body ground state), the quantum correlations are encoded in a fundamental object called the kernel, that is given by
\be
K_{M}\left(x,y\right)=\sum_{j=1}^{M}\psi_{j}^{*}\left(x\right)\psi_{j}\left(y\right),
\ee
where $\psi_1(x), \dots , \psi_M(x)$ are the wave functions of the $M$ single-body lowest energy states.
The quantum correlations of the fermions' positions can be expressed using the kernel. For instance, the density of particles at point $x$ is given by $\mathcal{R}_{M}\left(x\right)=K_{M}\left(x,x\right)$,  see e.g. Ref. \cite{DLMS16PRA} for more details.

The above expressions for $\mathcal{R}_{M}$ and $K_{M}$ are valid for arbitrary $M$, and (for the potential $V(x)= V_k(x)$) they describe the density and density-density correlations of the $M$ particles that remain near the interval $[-1,1]$, conditioned on a given value of the occupation fraction. However, in the large-$M$ limit they approach universal behaviors as we now describe.
Over large scales, the density is in general given by the local density (or Thomas-Fermi) approximation,
$\mathcal{R}_{M}\left(x\right)\simeq k_F\left(x\right)/\pi$
where
$k_F\left(x\right)\simeq\sqrt{2\left(\mu-V\left(x\right)\right)}$ is the local Fermi wave vector and $\mu$ is the Fermi energy (we consider $\hbar=m=1$). The density is normalized such that
$\int_{-\infty}^{\infty}\mathcal{R}_{M}\left(x\right)dx=M$.
For the square-well potential $V_k(x)$, as we saw above, $M$ is determined by the condition $\mu = E_M \simeq 0$, so we obtain
\be
\mathcal{R}_{M}\left(x\right)\simeq\begin{cases}
\sqrt{2k}/\pi \simeq M/2, & x\in\left[-1,1\right],\\[2mm]
0, & x\notin\left[-1,1\right],
\end{cases}
\ee
which thus determines $M \simeq 2\sqrt{2k}/\pi$ in agreement with Eq.~\eqref{approxM} above. This gives the mean inter-particle distance between particles inside the interval,
$d=1/\mathcal{R}_{M}\left(x\right) = 2/M$.
Moreover, the density correlations inside the interval are given in terms of the celebrated sine kernel \cite{eisler_prl, DLMS16PRA, DLMS2019review}
\be
K_{M}\left(x,y\right)\simeq\frac{\sin\left[k_{F}\left(x\right)\left(x-y\right)\right]}{\pi\left(x-y\right)}.
\ee
These expressions for the density and the kernel break down near the edges of the interval, $x = \pm 1$. In fact, the behavior of a noninteracting Fermi gas in the presence of a step-function potential was analyzed in Ref.~\cite{DLMSS21} in some detail. A universal form of the density and kernel were obtained. For $\mu\simeq0$ (the so-called critical case), they are given by
\be
\mathcal{R}_{M}\left(x\right)\simeq\frac{1}{d}n_{c}\left(\frac{x}{d}\right),\quad K_{M}\left(x,y\right)\simeq\frac{1}{d}\kappa_{c}\left(\frac{x}{d},\frac{y}{d}\right),
\ee
respectively, where the universal functions
$n_{c}$ and $\kappa_{c}$ are given explicitly in \cite{DLMSS21}. They provide an interpolation of $\mathcal{R}_{M}$ and $K_M$ between the regime inside the interval ($x \in [-1,1]$) and outside it ($x \notin [-1,1]$).

\section{Discussion} \label{sec5}

 We have studied the fluctuations of the occupation fraction of $N$ non-crossing Brownian particles in a finite interval.
 We found that the rate function that describes these fluctuations at long times undergoes a sequence of $N-1$ DPTs of second order. The $N$ different phases correspond to different numbers of particles in the vicinity of the interval.
 We achieved this by extending the DV formalism to noncrossing Brownian particles and found that this maps the problem to that of $N$ noninteracting spinless fermions in the presence of an effective potential. Under this mapping, the $N$ phases correspond to different numbers of bound single-body states of the effective potential.

  This DPT observed here for any $N \geq 2$ is of very different nature to the DPT that has been previously found for a single particle ($N=1$). The latter DPT only occurs if the particle experiences an additional external drift (which leads to breaking of time-reversal symmetry)
 \cite{TouchetteMinimalModel, Touchetteoneparticle}, and its properties are quite different to those of the DPTs that we found here. For instance, it involves coexistence between two different phases, 
 which correspond to the particle being confined (or not) to the vicinity of the interval. 
 However, in the present work, we have found $N-1$ DPTs that separate between $N$ temporally-homogeneous phases, each corresponding to a different number of particles which are in the vicinity of the given interval for the entire dynamics.

  It would be interesting to study fluctuations of the occupation time in presence of other interactions between the Brownian particles (instead of or in addition to the non-crossing condition). 
   This problem has been solved in the limit $N\gg1$ for a broad class of interacting diffusive gases (e.g., the exclusion process), that does not include the gas of noncrossing Brownian particles as a particular case, in \cite{AKM19}. 
  Another natural extension of the present problem would be to study the effect of a drift in the system, or to study the occupation time for other processes such as fractional Brownian motion or for active particles.
  Moreover, one could study more detailed quantities such as the empirical distribution of the positions of the particles.


\subsection*{Acknowledgments}

We acknowledge several useful discussions with Pierre Le Doussal on related topics.

\appendix

\medskip

\section{Donsker-Varadhan formalism for noncrossing particles}
\label{feynman}

We first consider the unnormalized joint probability distribution $Q(X_1, \dots, X_N, \mathcal{R}, t)$  of the $X_i$'s and $\mathcal{R} = t \rho_t$. $Q$ satisfies the Fokker-Planck equation
\be
\frac{\partial Q}{\partial t}= L^\dagger \,  Q
-  \mathds{1}_{[ -1,1 ]}(x) \, \frac{\partial Q}{\partial \mathcal{R}}
\ee
together with the non-crossing boundary conditions given in Eq. \eqref{eqbc} at $X_i = X_{i+1}$.
It is convenient to define the generating function 
\be \label{kac2}
G_k(X_1, \dots, X_N, t) =  \int e^{k \,  \mathcal{R}} Q(X_1, \dots, X_N, \mathcal{R}, t) d \mathcal{R}.
\ee
According to Feynman-Kac formula \cite{kac, majumdar5},  $G_k$ evolves with time as
\be
\label{dGkdt}
\frac{\partial G_k(X_1, \dots, X_N, t)}{\partial t} =  \mathcal{L}_k  \,  G_k(X_1, \dots, X_N, t)
\ee
where 
 \be
 \mathcal{L}_k  = L + k \mathds{1}_{[ -1,1 ]}(x)
\ee
known as the tilted generator.  Note that $G_k$ also satisfies the boundary conditions \eqref{eqbc}.

Eq.~\eqref{dGkdt} can be solved by expanding $G_k$ over the eigenbasis of $\mathcal{L}_k$
\cite{footnote:SumIntegral}
\be \label{kac3}
G_k(X_1, \dots, X_N, t) = \sum_i \, c_i \,  e^{t \zeta_i} \,  r_k^i (X_1, \dots, X_N)
\ee
where $\zeta_i$ and $r_k^i (X_1, \dots, X_N)$ are the eigenvalues and eigenfunctions of $\mathcal{L}_k$
respectively, and the $c_i$'s are coefficients that depend on the initial condition.

For $T \longrightarrow \infty$, the sum in Eq. \eqref{kac3} is dominated by the largest eigenvalue $\zeta_{\max}$ (and if $\zeta_{\max}$ doesn't exist then instead of it we must use the supremum of the $\zeta_i$'s).
In this long time limit the nomalization factor $\mathcal{N}(t)$ goes as a power law  $\mathcal{N}(t) \sim t^{- N (N-1)/4}$  \cite{fishersurvival, fishersurvival1, viciouswalker, viciouswalker1, GLMS19}. Comparing Eq. \eqref{kac3}  to the definition of the SCGF in Eq.~\eqref{eqscgf} it can be concluded that $\lambda(k) = \zeta_{\max}$, as the logarithm of $\mathcal{N}(t)$ can be neglected. 


\section{Asymptotic behaviors of $\lambda(k)$ and of $I(\rho)$} \label{assymp calculation}

In this section, we show the details of the calculation of some of the asymptotic behaviors of the SCGF and the rate function. We mostly focus on the case $N=2$, and consider the limits $k \longrightarrow 0$ ($\rho \longrightarrow 0$), $k\rightarrow\infty$ ($\rho \longrightarrow 2$) and  near the critical point, $k\simeq k_c$. We use Eqs.~\eqref{eq3}-\eqref{eq31new} to study the asymtotic behaviours for different limits of $A$ (or $k$).

\subsection{$k \simeq k_c$}

For any general $N$, for $k$, that is near the critical point $k_{c,i} =  \frac{i^2\pi^2}{8}$ but slightly larger (i.e., for $k>k_{c,i}$ with $k - k_{c,i} \ll k_{c,i}$),  using  $\tan x=\tan\left(x-n\pi\right)\simeq x-n\pi$ (for even $i$) and $\cot x=\cot\left(x-n\pi\right)\simeq x-n\pi$ (for odd $i$) for  $x\simeq n\pi$, the $k$ and $\lambda_{i+1}(k)$ behaves for $A_i \longrightarrow \frac{i^2\pi^{2}}{8}$ as 
  Eq.~\eqref{eq3} becomes
\be
k \simeq \frac{1}{2}\left(\frac{i^2\pi^{2}}{8} - A_i\right)^{2}+ A_i
\ee
leading to
\bea
\lambda_{i+1}(k) &=& k-A_i \simeq \frac{1}{2}\left(\frac{i^2\pi^{2}}{8} - A_i\right)^{2} \nn\\
&\simeq& \frac{1}{2}\left(k-\frac{i^2 \pi^{2}}{8}\right)^{2}
\eea
which coincides with Eq. \eqref{asymscgfgeneral} of the main text. Using the above two equations the behaviour of the SCGF $\lambda(k)$ for $N=2$ can now be written as
\be 
\lambda(k) \simeq  \lambda_1(k) + \frac{1}{2}\left(k-k_c\right)^{2}
\ee
which is Eq. \eqref{eqscgfcritical} in Sec. \ref{sec4}. 
Taking the derivative we have:
\be
\label{rhoOfknearkc}
\rho=\frac{d\lambda}{dk}\simeq\lambda_{1}'(k)+(k-k_{c})\simeq\rho_{c}+\left[\lambda_{1}''(k_c)+1\right](k-k_{c})
\ee
where we used 
$\rho_{c}=\lambda_{1}'(k_{c})$.
Inverting this relation, we obtain
\be
\label{kOfrhonearkc}
k \simeq k_c + \frac{\rho-\rho_{c}}{\lambda_{1}''(k_c)+1}.
\ee
Taking the Legendre transform of the Eq.~\eqref{eqscgfcritical} while using Eqs.~\eqref{rhoOfknearkc} and \eqref{kOfrhonearkc}, the expression of the rate function at $\rho$ that is slightly larger than $\rho_c$ can now be expressed as
\be \label{ratefunctioncritical}
I(\rho) \simeq I_{1}(\rho_{c})+k_{c}(\rho-\rho_{c})+\frac{1 }{2\left[\lambda_{1}''(k_c)+1\right]} \left(\rho-\rho_{c}\right)^{2}
\ee
which coincides with the second line of Eq.~\eqref{eqratefunccritical} in Sec. \ref{sec4}.


\subsection{$k \to 0$}

In the $k\rightarrow 0$ limit, the SCGF $\lambda(k) = \lambda_1(k) \approx 0$, which implies $A \longrightarrow 0 $. Around $A=0$, using Eqs.~\eqref{eq3} and \eqref{eq31new}, the asymptotic behaviors of $\lambda(k)$ and $k$ are
\be
\lambda(k) \simeq 2A^{2},\qquad k\simeq A+2A^{2}
\ee
respectively. This implies
\be
\lambda(k) \simeq 2k^2,
\ee
coinciding with the first line of Eq. \eqref{inftyasymscgf} in Sec. \ref{sec4}. The Legendre transform of $\lambda(k)$ in this limit is thus
\be
I(\rho) \simeq \frac{\rho^{2}}{8}
\ee
which coincides the first line of Eq. \eqref{inftyasymrate2} in Sec. \ref{sec4}. 

\subsection{$k \to \infty$}

In the limit  $k \longrightarrow \infty$,  $A_1$ goes as  $A_{1}\rightarrow\frac{\pi^{2}}{8}$. Considering Eq. \eqref{eq3} the asymptotic behavior of the left hand side of Eq.~\eqref{eq3} near this value of $A_1$ we obtain
\bea
\label{kOfA1approx}
 k &\simeq&   \frac{\pi^{4}}{32\left(\frac{\pi^{2}}{8} - A_1\right)^{2}} - \frac{3\pi^{2}}{8\left(\frac{\pi^{2}}{8} - A_1 \right)}+\frac{1}{24}\left(21+\pi^{2}\right)\nonumber \\ 
 &-&\left(\frac{1}{3}-\frac{1}{2\pi^{2}}\right)\left(\frac{\pi^{2}}{8} - A_1\right)
 \eea
which yields
\bea
\label{lambda1OfA1approx}
\lambda_1(k)  &=& k - A_1  \nn\\
&\simeq& \frac{\pi^{4}}{32\left(\frac{\pi^{2}}{8} - A_1\right)^{2}} - \frac{3\pi^{2}}{8\left(\frac{\pi^{2}}{8} - A_1\right)}-\frac{\pi^{2}}{12}+\frac{7}{8}\nonumber \\
 &+&\left(\frac{2}{3}+\frac{1}{2\pi^{2}}\right)\left(\frac{\pi^{2}}{8} - A_1\right).
 \eea
 From Eqs.~\eqref{kOfA1approx} and \eqref{lambda1OfA1approx}, we now express $\lambda_1(k)$ as a function of $k$ in the limit $k \gg 1$
 \be
 \label{lambda1Ofkapprox}
 \lambda_1(k) \simeq k-\frac{\pi^{2}}{8}+\frac{\pi^{2}}{4\sqrt{2k}},
 \ee
  which coincides with the second line of Eq. \eqref{inftyasymscgf} in the main text.
This result gives the asymptotic behavior of the SCGF for the case $N=1$. 

For $N=2$, we must add the contribution of $\lambda_2(k)$, whose asymptotic behavior in the large-$k$ limit we now calculate.
For the second energy level, $A_2$ goes as $A_{2}\rightarrow\frac{\pi^{2}}{2}$ and  by using the corresponding behavior of Eq.~\eqref{eq3new} and following similar steps as above, we find
 \bea
 \label{lambda2Ofkapprox}
\lambda_2(k) \simeq k-\frac{\pi^{2}}{2}+\frac{\pi^{2}}{\sqrt{2k}} \, .
 \eea
 Now using Eqs.~\eqref{lambda1Ofkapprox} and \eqref{lambda2Ofkapprox}, we find the asymptotic behavior of the SCGF $\lambda(k) = \lambda_1(k) + \lambda_2(k)$ at $k \longrightarrow \infty$
 \be
 \label{lambda12Ofkapprox}
\lambda(k) \simeq 2k-\frac{5\pi^{2}}{8}+\frac{5\pi^{2}}{4}\frac{1}{\sqrt{2k}}
\ee
 which coincides with Eq. \eqref{inftyasymscgf2} in Sec. \ref{sec4}. 
 We now calculate the Legendre transform of this expression, in order to find the corresponding asymptotic behavior of the rate function.
 Taking the derivative of Eq.~\eqref{lambda12Ofkapprox} w.r.t $k$, we get the limiting behaviour of $\rho \longrightarrow 2$ as
\be
\label{rhoOfkapprox}
\rho=\frac{d\lambda(k)}{dk}=2-\frac{5\pi^{2}}{8\sqrt{2}k^{3/2}}.
\ee
By combining Eqs.~\eqref{lambda12Ofkapprox} and \eqref{rhoOfkapprox},
we find the behavior of the rate function in the limit $\rho \longrightarrow 2$,
\be
I(\rho) = k\rho - \lambda(k) \simeq
\frac{5\pi^{2}}{8}-3\left(\frac{5\pi^{2}}{8\sqrt{2}}\right)^{2/3}\left(2-\rho\right)^{1/3},
\ee
which coincides with Eq. \eqref{inftyasympI} in Sec. \ref{sec4}.
 The second line of Eq.~\eqref{inftyasymrate2} (for the case $N=1$) is obtained similarly, by calculating the Legendre transform of Eq.~\eqref{lambda1Ofkapprox}.



\begin{thebibliography} {99}
\bibitem{footnote:noncrossing}
We emphasize that the process that we consider is $N$ Brownian motions conditioned not to cross each other. This is not be confused with a hard-core repulsion between the particles. The latter would result in reflecting boundary conditions, instead of the absorbing boundary conditions \eqref{eqbc} for the noncrossing case.
\bibitem{footnote:SlaterNormalization} The absence of the usual normalization factor $1/\sqrt{N!}$ before the Slater determinant in Eq.~\eqref{SlaterDet} is because it is constrained to $X_1 < \dots < X_N$ in contrast to the usual case with fermions, thus leading to a different normalization factor.
\bibitem{footnote:SumIntegral} The $\sum$ in Eq.~\eqref{kac3} is to be understood as a sum over the discrete part of the spectrum of the tilted operator plus an integral over the continuous part of the spectrum.
\end{thebibliography}
\end{document}